# An optimized hybrid solution for IoT based lifestyle disease classification using stress data

SADHANA TIWARI[1], SONALI AGARWAL[2]

*Indian Institute of Information Technology, Allahabad, India* [1, 2]

*rsi2018507@iiita.ac.in*[1]*, sonali@iiita.ac.in*[2]

**Abstract**

Stress, anxiety, and nervousness are all high-risk health states in everyday life. Previously, stress levels were determined by speaking with people and gaining insight into what they had experienced recently or in the past. Typically, stress is caused by an incidence that occurred a long time ago, but sometimes it is triggered by unknown factors. This is a challenging and complex task, but recent research advances have provided numerous opportunities to automate it. The fundamental features of most of these techniques are electro dermal activity (EDA) and heart rate values (HRV). We utilised an accelerometer to measure body motions to solve this challenge. The proposed novel method employs a test that measures a subject's electrocardiogram (ECG), galvanic skin values (GSV), HRV values, and body movements in order to provide a low-cost and time-saving solution for detecting stress lifestyle disease (i.e. diseases caused by a person's lifestyle or habits, such as stress, obesity, and so on) in modern times using cyber physical systems. This study provides a new hybrid model for lifestyle disease classification that decreases execution time while picking the best collection of characteristics and increases classification accuracy. The developed approach is capable of dealing with the class imbalance problem by using WESAD (wearable stress and affect dataset) dataset. The new model uses the Grid search (GS) method to select an optimised set of hyper parameters, and it uses a combination of the Correlation coefficient based Recursive feature elimination (CoC-RFE) method for optimal feature selection and gradient boosting as an estimator to classify the dataset, which achieves high accuracy and helps to provide smart, accurate, and high-quality healthcare systems. To demonstrate the validity and utility of the proposed methodology, its performance is compared to those of other well-established machine learning models.

## 1. Introduction-

Stress has become an increasingly widespread ailment as the modern world has gotten faster and faster. Stress levels in India are substantially greater than in many other countries; around 82 percent of Indians feel stressed [1]. The 35-49 year old age group is the most stressed. Stress is defined as a reaction to psychological, physiological, behavioural, and situational restrictions in combination. It's a stimulation state in which things run out of control and become extremely difficult. Regular stress can have a number of detrimental health repercussions [2]. Extreme levels of stress may be harmful to people's health and well-being, resulting in lack of sleep, high blood pressure (hypertension), depression, mental illnesses, and heart difficulties, among other things. Acute stress is temporary; but, when the degree of stress rises, it may develop into chronic stress, which is difficult to alleviate or even manage. These negative consequences impair an individual's health and well-being, as well as diminish overall efficiency, productivity, and peace of mind [3][4].

A cyber physical system (CPS) is a computer system in which computer-based algorithms operate or monitor a system. Physical and software components are closely linked in CPS, allowing them to function on many spatial and temporal dimensions, show diverse and unique behavioural modalities, and interact with one another in context-dependent ways [5]. Cloud computing and big data technologies are also critical in the development of the Health-CPS patient-centric healthcare system. Data acquired by cyber-physical health systems may be successfully sent to cloud storage [6]. Machine learning (ML) models are employed with CPS to enhance the quality of healthcare services and analyse data to assist healthcare workers in making choices [7]. Smart grids, autonomous automotive systems, medical monitoring, industrial control systems, robotics systems, and automated pilot avionics are all examples of CPS. Computation, networking, and physical processes are all integrated in Cyber-Physical Systems (CPS). Physical processes are monitored and controlled by embedded computers and networks, which include feedback loops where physical activities effect calculations and vice versa. By offering abstractions and modelling, design, and analytic approaches for integrating the entire system, CPS connects the dynamics of physical processes with software and networking [8][9].

Lifestyle illnesses are caused by a person's lifestyle or habits, or how they live their lives. There are various factors that have contributed to the fast increase of these diseases in recent years, including poor eating (junk food), lack of exercise, and addictive hazardous habits such as smoking and drinking. These illnesses are non-communicable by their very nature of spreading [10]. These diseases are more likely to develop in the body if you live an irregular and unhealthy lifestyle. Long-term activity monitoring can aid in the treatment of lifestyle-related disorders such as stress, obesity, diabetes, and hypertension [11]. Frequent monitoring of many activities and physiological markers of the human body, including EDA, can also aid in the early detection of several chronic lifestyle disorders, including stress. IoT sensors, including as ECG, Galvanic skin response (GSR), Oxygen saturation (SPO2), and Accelerometer sensors, play a critical role in precise monitoring of physiological body factors connected to stress [12]. Monitoring environmental elements that might cause or treat any ailment, as well as monitoring other everyday activities and physiological indicators, are all big uses of IoT in the medical arena that have piqued the interest of researchers and medical experts [13]. This study proposes a novel low-cost and time-efficient machine learning approach for analysing an individual's stress. The following are the significant contributions of this study:

- Computation of a stress score in response to a questionnaire. The subject is not stressed if the stress score is less than the threshold number, and vice versa.

- If the individual is stressed, we will examine the person's physiological and psychological factors using the HR, GSV, ECG, and accelerometer

- Using the Synthetic Minority Oversampling approach (SMOTE) to handle the class imbalance problem to prevent biased findings

- Using the Grid Search approach to optimise the model by selecting the optimum set of parameters.

- Introducing a novel CoC-RFE approach, which runs the feature selection algorithm to extract the most valuable characteristics and helps to remove the irrelevant features followed by execution of various machine learning methods to calculate the accuracy of the models

- A hybrid classification model (CoC-RFE-GB) is proposed for better accuracy and performance.

The primary goal of this study is to provide an accurate and practical solution for smart healthcare monitoring based on stress data leveraging internet of things (IoT). The paper is organized as follows: The introductory details are explained in Section I, and the reviewed relevant work in this field is discussed in Section II. The methodology of the work is described in Section III. In section IV, we performed a comparative analysis of classification results of each model based on statistical factors. Section V focuses on the in-depth graphical representation of the results. The last section VI contains concluding observations as well as recommendations for further research.

## 2. Related work

To find the different parameters, methodologies, and models employed by the researchers, a thorough review of previous work was conducted. According to the Cigna 360 Happiness Survey 2019, India's stress levels are still very high when compared to other developed and developing countries such as the United States, the United Kingdom, Germany, France, and Australia, and the sandwich generation, aged 35 to 49, is the most stressed, followed by the millennial generation. Approximately 89 percent of oppressed respondents were somewhat worried, compared to 87 percent of the millennial generation and 64 percent of the population over 50 years. In contrast to the overall findings, 85 percent of Indian males and 82 percent of working women reported greater levels of stress. Overwork and financial concerns about personal health are the primary drivers of stress for women [1]. A research work proposed, ANN based hybrid model for stress and mental health detection of students aged between 18-35 uses GSR and HR values and achieves 99.4% accuracy. This work also predicts the stress levels of each subjects into four different classes [3]. Feng-Tso Sun et al. offer an activity-based mental stress inquiry approach based on continuous monitoring of vital body parameters related to stress employing ECG, GSR, and Accelerometer sensors. Using the feature derived by the accelerometer, as well as the GSR and ECG sensors, this article illustrates the influence of physical activities. Twenty people were chosen to participate in a set of activities meant to assess a person's stress level. Because the heart rate fluctuates fast between activities, the findings were better when no ECG data was used. Later with inclusion of GSR sensor, the decision tree classifier with all features had the best classification accuracy (92.4 %) [4]. A model proposed by Tiwari, Sadhana, et al collects data through IoT sensors such as GSR and ECG for recognition of human emotion. It classifies the emotions into four classes i.e. anger, happy, sad and joy using ANN [12]. Stress has been linked to an increased risk of heart disease in several studies. A new stress sensor is designed using GSR which is controlled through Zigbee. 16 persons (8 men and 8 women) were chosen to participate in an experiment employing this newly built sensor. The GSR sensor is capable of recognising distinct emotional states of each user once the trial is completed, with a success rate of 76.56 % [14].

Fitri Indra Indikawati et al. developed a customised stress inquiry system based on data obtained from IoT sensors in another study. The WESAD dataset is used to classify stress in this system. For the four classes of stress, amusement, baseline, and meditation, three machine learning classification models were used: Logistic Regression, Decision Tree, and Random Forest. After comparing the performance of the developed model with the performance of several classifiers, the random forest classifier provides the most accurate and consistent stress recognition, with accuracy ranging from 88% to 99% for 15 items [15]. Jennifer A. Healey suggested a new dataset based on physiological signals that includes variations in various body parameters from day to day. The original data was collected by evaluating young individuals in stressful situations such as rush hour or driving on highways, and numerous parameters were recorded, including ECG, EMG, HR, GSR, etc.. The model for identifying a user's emotion level was created using the Python automated machine learning packages TPOT. The characteristics discovered in this unique dataset have an 81% success rate for all eight types of emotions and a 100% success rate for subsets of various emotion-based attributes [16]. Mario Salai et al. describe the findings of two studies done with the use of a low-cost heart rate sensor and a chest monitor for stress detection. The trial included 46 healthy volunteers, the majority of whom were students (27 men and 19 women; average age: 24.6 years). The experiment was split into two portions, each lasting 10 minutes, for a total of 20 minutes. The mean HR, pNN50, and RMSSD characteristics are combined to indicate stress, and a brute force technique is employed to establish the HRV feature's threshold. The rest state is not detectable by this technique. For accurately identifying the stress, the accuracy, sensitivity, and specificity scores are 74.60 %, 75%, and 74.19 %, respectively [17].

Alberto Greco et al. presented a unique approach for electrodermal activity (EDA) analysis, which consists of changes in electrical characteristics of skin such as resistance, conductance, and voltage by utilising convex optimization techniques [18]. Another study found that using data collected by various wearable sensors, various machine learning and deep learning approaches are highly beneficial in the stress detection process [19]. K-Nearest Neighbour, Linear Discriminant Analysis, Random Forest, Decision Tree, AdaBoost, Gradient Boosting, and Kernel Support Vector Machine are some well-known models used for stress detection. Furthermore, a simple feed forward deep learning artificial neural network exhibiting high accuracy [19] [20]. The major goal of this study is to develop a simple way for detecting stress levels in people, so that they may be alerted if their stress levels are excessive.

## 2.1 Challenges and Research Gap

The process of data extraction and pre-processing of the raw file of the WESAD dataset is a very tedious task for this problem statement. Another difficult challenge is to identify the most significant set of features from the dataset, as it contains many features. Some of the features are not really that useful in the stress proposed study. After getting the optimal set of features, we can reduce the execution time of the model. Monitoring a person's physical activity is also vital for improving the stress detection process. These characteristics provide information on a participant's everyday routine. Apart from that, the stress score is calculated using a questionnaire. Till now, we have been able to determine the stress level using the EDA and heart rate. Then following study questions may arise:
1. Have you experienced a high heart rate while doing any heavy workout or physical activity?
2. Have you experienced high sweating while doing some physical activity or heavy activity?

The answer to both questions is yes, and the model provided is paradoxical (since this model will anticipate stress while you are working out when you are not anxious). We deployed an accelerometer to measure body motions to solve this challenge. A test kit that measures the ECG, GSV, and HRV values is used in the present approach. These are recorded, and a machine learning algorithm is used to determine whether or not a person is stressed. Because each individual must wear the gadgets and the data must be recorded, this is a costly and time-consuming operation. A new system has been introduced to minimise expenses and save time, in which each subject have two options:

1. Get an ECG, GSV, or HRV test to get more reliable findings.
2. Use a questionnaire to calculate stress levels, which will yield somewhat less accurate findings.

If a subject voluntarily chooses option 1, the ECG/GSV/HRV values will be recorded by showing him comedy videos and mental arithmetic exercises. After that, they could ask to fill a questionnaire. This would be a simple stress detection questionnaire and it would have options. The responses given by the subject are recorded. Since the stress level of a subject is known and identified as a label and features are identified as the options. Then the model is trained to give scores to each option. If a subject opts for 2nd, the questionnaire is given to him and the stress score is provided according to the responses he makes. Since data collection was not possible therefore we just trained our model on the 70% of the WESAD dataset and tried to predict the stress level on the rest 30% of the dataset. The accuracy achieved around 98.88 % on the overall dataset and a mean accuracy of 99% on doing individually on each subject.

## 3. Material and methods-

This section goes into the materials and procedures needed to conduct the experiment and calculate the results. There are six sections to this part. The first section explains the WESAD dataset, which is a publicly accessible multimodal time-series dataset. The experimental setup is described in the second section. Part three discusses the difficulties that come with parameter selection, data cleaning, and data pre-processing. The suggested classification model's setup and execution, as well as the statistical metrics utilised to validate the proposed model, are covered in the fourth section.

### 3.1 Dataset description

The proposed method uses WESAD dataset [15] [21], which is a publicly available dataset for wearable stress and affect detection. The data is recorded from 15 subjects using a chest (RespiBAN) and wrist (Empatica E4) worn devices. These sensors are used to record blood volume pulse, electrocardiogram, electro dermal activity, electromyogram, respiration, body temperature, and three-axis acceleration. Device Empatica E4 records the blood volume pulse (BVP), body temperature, electro dermal activity (EDA) etc. and device RespiBAN records electrocardiogram (ECG), EDA, respiration, temperature and 3-axis acceleration.

### 3.2 Hardware setup used for Data acquisition

The data is collected using two very significant IoT devices: chest wearable device (RespiBAN) and wrist wearable device (Empatica E4), then 29 features are selected, which are related to BVP, EDA, temperature, respiration and their minimum, maximum and mean values. Apart from these, age and weight are also taken as features. The stress labels assigned across these data are taken as 0, 1, and 2 [15].

### 3.3. Data preparation

Data preparation includes the challenges associated in selection of various significant input parameters and data pre-processing phase.

### 3.3.1 Selecting input parameters

The frequent metrics discovered for predicting stress after evaluating several research publications are EDA, GSR, HRV, ECG, and others.

(i) I EDA (Electrodermal Activity)- According to the conventional EDA hypothesis, skin resistance fluctuates depending on the status of sweat glands in the skin. The sympathetic nervous system (SNS) regulates sweating, and skin conductance is a sign of psychological or physiological arousal. Sweat gland activity rises when the sympathetic branch of the autonomic nervous system (ANS) is significantly stimulated, which raises skin conductance. Skin conductance can thus be used to measure emotional and sympathetic reactions [22]. GSR (galvanic skin response) is a measure that describes how sweat gland activity changes in response to changes in our emotional state [19]. The GSR is used to track variations in sweat gland activity or skin conductance over time. There is a link between GSR and stress levels, according to several research. The larger the conductance, lower the resistance, stronger sweat organ activity, and higher the value of GSR, the more stressed a person is [23].

(ii) HRV (heart rate variability) - This is a measurement of the time difference between successive heartbeats, as shown in Figure 1. The variance between subsequent heartbeats is low when a person is agitated or in fight-or-flight mode, i.e. HRV is low in panic circumstances. HRV, on the other hand, is high during relaxation, suggesting a healthy nervous system; a high HRV, on the other hand, shows balance. People with a low HRV value are more likely to be stressed, whereas those with a high HRV are more likely to be healthy [14].

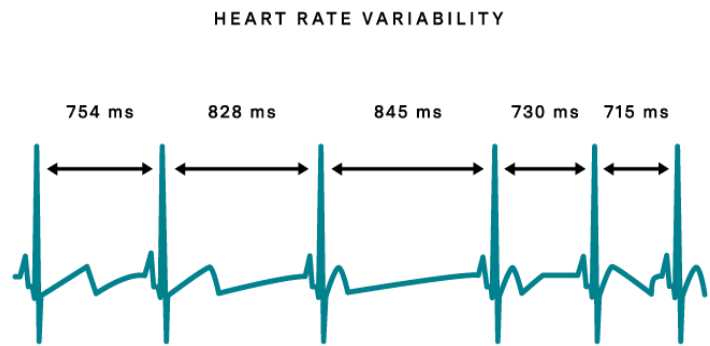

Fig. 1: Heart Rate Variability [8]

(iii) ECG (Electrocardiography) - It is used to measure the electrical activity of the heart and its health. It detects various heart related issues like irregular heart rhythms. ECG generates a graph as output where time is on x-axis and voltage (amplitude) on y-axis, usually the readings of the graph is divided in squares where,
1 big square consist of 25 mini squares
1 big square =0.5mV (on y-axis)
1 big square =0.20 seconds
1. When there is regular rhythm in QRS complexes (peak) then
    Heart rate=300 No. of large squares between two qrs complexes
2. When rhythm is irregular heart rate is given as
    Heart rate=10(number of R values (peak) in 6 sec.)

Some other features can also be considered for the stress detection experiment,; however, we must evaluate which features are most significant for our investigation. One of the most essential aspects used to identify stress is the galvanic skin response (GSR). In a study done by Villarejo et al., they were able to reach 76.56 % by utilising solely GSR characteristics [14][23].

### 3.3.2 Data Pre-processing

Data pre-processing is an important part of data mining since it aids in the transformation of raw data into something that is relevant, usable, and well-organized. One of the most important phases in dealing with duplication, irrelevant, missing, and noisy data is data cleansing. Individual data files are combined into a single merged file via data integration [24] [25]. Each subject's data is saved in pickle files. This file extracts the information and transforms it to a standard csv file with all of the necessary characteristics. The algorithmic methods used to clean data are as follows:

Step 1: A class has been created, which will have attributes related to the properties of a subject (signal, label and subject), types of signals (chest and wrist), data recorded by chest device (ACC, EMG, EDA, Temp, etc.) and data recorded by wrist device (ACC, BVP, EDA, TEMP).
Step 2: The constructed class has member functions which returns the wrist data and chest data and a function which performs mapping using a key for the type of data recorded in the wrist device and value as the variance of the data recorded.
Step 3: The data from the pickle files is loaded for each subject.
Step 4: For each data, the minimum, maximum and standard deviation value for BVP, EDA (phasic), EDA (smna), and temperature is calculated. Also the age and weight of the subject are extracted from the pickle file.
Step 5: BVP peak frequency is calculated using the built in period gram function in scipy.signal module.
Step 6: These data are then pushed into the data frame and data of all subjects is merged into one.
Step 7: The final data frame is written into the m14_merged.csv. This is the cleaned dataset file which is used for classification purposes.

The next step involves concatenating new EDA features therefore Electrodermal Activity Processing is done by using a Convex Optimization Approach. This method generates phasic, tonic, and additive white Gaussian noise, which takes into account model prediction errors, measurement errors, and artefacts [18].

### 3.4 Implementation of the proposed model

The proposed model's implementation has been broken down into two primary parts, namely Pre-processing the raw data, dealing with data class imbalance concerns, and conducting feature selection on different sample sizes to produce an ideal feature subset and discover the best feature reduction strategies for the processed data are all part of the first stage. The feature vector is then classified using a fine-tuned set of hyper parameters for the model in the next stage. The data is captured in its raw form, and it must be pre-processed before our classifier can be trained on it. These two phases might then be subdivided into four parts:

### 3.4.1 Handle Class Imbalance problem

Class imbalance is one of the major issues that should be addressed carefully and efficiently during classification. Assume that if we are dealing with a binary class classification problem, where the bulk of the samples belong to one class and only a few examples belong to the other. As a result, if we perform classification using class imbalance data, the results may be skewed in favour of the majority class, because the majority class's contribution to the classification model will be greater than the minority class's. As a result, the model's performance will be affected. As a result, we adopted the SMOTE technique introduced by Chawla et al. in 2002 [26] [27] to deal with the problem of class imbalance. The core idea behind this approach 's to keep the class balanced by producing synthetic samples from the minority class. The synthetic random samples are generated using the k-nearest neighbour approach.

Algorithm-I
Pseudocode of SMOTE

SMOTE (T, N, k)
Input:
Input Data $D_i = \{x_i \in X, \forall i = 1, 2,......T_m\}$
Minority target samples $T_m$;
Quantity of SMOTE P%;
Nearest neighbour's 'k.'
 Output: (P/100) *$T_m$ = $T_s$ synthetic minority class samples
 Algorithm:
For i = 1,2,......,$T_m$ do
1.   Select k-nearest neighbour's from $T_m$ of $x_i$
2.    P' = P/100
While P'! = 0 do
(Perform step 3 to 7)
3.   Choose any of the k-nearest neighbour's x'
4.   A random number $\delta$ must be chosen from the interval [0,1]
5.   $x_s = x_i \cup \delta*(x' - x_i)$
6.   adjoin $x_s$ to $T_s$
7.   Decrease P' by 1
End while

**3.4.2 Feature reduction and selection of optimal subset of features**

By finding the irrelevant and undesired characteristics, feature reduction helps to reduce the number of attributes in a huge data collection. This is essentially a data pre-processing step for large volumes of data. As illustrated in figure 2, the notion of feature selection indicates that we pick a subset of features or characteristics that assist in the creation of an efficient model to represent the selected subset of features [28].

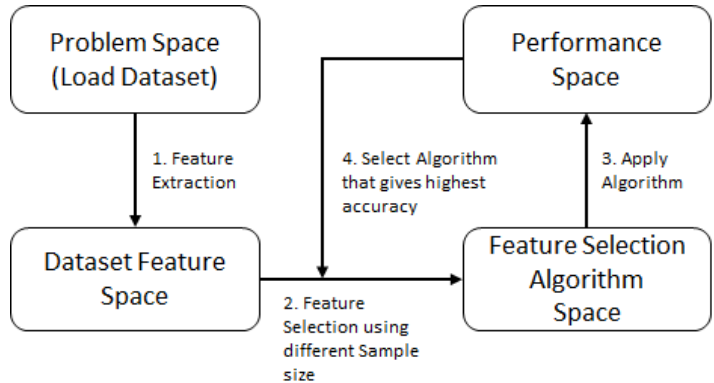

Fig. 2: Process of Optimal feature selection

By lowering the quantity of the input variables and identifying the ideal collection of features, the model's performance may be enhanced. Additionally, it is essential to reduce the computational cost and complexity of the model, as well as to improve prediction outcomes caused by overfitting. Apart from selecting the ideal feature subset, it also accomplishes a number of other goals, including dimensionality reduction, condensing the quantity of data required for the learning process, and increasing the efficiency of the generated model [29]. Four well-known feature selection strategies have been employed in this work: ANOVA f-test, Mutual Information (MI), Recursive Feature Elimination (REF), and Correlation coefficient. We examined all of these approaches using a variety of sample sizes, including 10, 20, 30, 40, and 50, and found that a sample size of 40 produces the best subset using all four strategies.

Algorithm-II
Pseudocode of correlation based RFE (CoC-RFE) feature selection algorithm

```
CoC-RFE (D, N, F, n)
Input:
D: Input Data
Dtrain: Training data
F: Original set of features, F= [f1, f2, f3,…………,fn]
d: Dimension of feature
N: Size of population
C(Dtrain, F): Feature ranking technique
Output: SF: Optimal subset of features
Algorithm:
1.   Initialize instance of the population
         S_F = F
2.   Compute correlation coefficient matrix R using features in F
3.   Find out the contribution of each feature of F with the help of R
4.   If correlation coefficient threshold (Cf) < Threshold (Cλ) then
5.     Discard the feature fi from Fn
6.   Else
7.   Add the feature in SF
8.   while termination condition (Cf < Cλ) is not true do
9.     for j = 1 to n do
10.         Rank the feature S_F by applying C(Dtrain, S_F)
11.         f_1 = Least important feature in S_F
12.         Update the best position of the feature using Rank of features in S_F
13.         S_F = S_F - f_1
14.   end for
15.   end while
```

### 3.4.3 Hyper parameter selection for selected features

Hyper parameters are very important because they determine the overall behaviour of a machine learning model. The main goal of parameter tuning is to find the optimal set of hyper parameters which minimizes the predetermined loss function to get the best results. This is also necessary to avoid the issues of overfitting and under fitting as no single value of parameters well suits all ML models. In this work, the Grid search (GS) method is used for hyper parameter tuning [30] [31]. All the well-known machine learning models used in this experiment are trained by applying training data using a selected set of hyper parameters across each model. After that, evaluation of the model is done using the test data with the help of some statistical measures such as accuracy, f1-score, recall and precision etc. The fine-tuned set of hyper parameters used for various machine learning models used in this experiment have been listed in table 1.

Table 1. Hyper parameters used across various machine learning models

| ML Classifier | Hyper parameters | Values used for selection of best hyper parameters | Best hyper parameter used |
|---|---|---|---|
| LR  | solvers | ['newton-cg', 'lbfgs', 'liblinear'] | ['newton-cg'] |
|     | penalty | ['l2'] | ['l2'] |
|     | c_values | [200, 100, 10, 1.0, 0.1, 0.01] | [100] |
| SVC | kernel | ['poly', 'rbf', 'sigmoid'] | ['rbf'] |
|     | C | [100, 50, 20, 10, 1.0, 0.1, 0.01] | [50] |
|     | gamma | ['scale'] | ['scale'] |
| KNN | n_neighbors | Range(1, 21, 2) | [1] |
|     | weights | ['uniform', 'distance'] | ['uniform'] |
|     | metric | ['euclidean', 'manhattan', 'minkowski'] | ['manhattan'] |
| LDA | 'solver' | ['svd', 'lsqr', 'eigen'] | ['svd'] |
| RF  | estimators | [10, 50, 100, 150, 500, 700, 1000] | [1000] |
|     | max_features | ['log2', 'sqrt'] | ['log2'] |
| GB  | n_estimators | [10, 100, 500, 1000] | [1000] |
|     | learning_rate | [0.001, 0.01, 0.1, 0.5] | [0.1] |
|     | subsample | [0.5, 0.7, 1.0] | [0.7] |
|     | max_depth | [1, 3, 7, 9, ] | [3] |

### 3.4.4 Model setup of the proposed model

This section develops a novel hybrid model, CoC-RFE-GB, that outperforms current machine learning models in terms of accuracy. Figure 3 depicts the planned CoC-RFE-GB system's work-flow diagram. This innovative classification model outperforms all other well-known machine learning classification models in terms of accuracy. The following phases comprise the systematic processing of the newly developed hybrid model:

**STEP -I -** The raw data is taken as input for the designed model. The WESAD dataset cannot be used for classification in its original form, thus we need to pre-process it. WESAD contains separate directories and sub directories for each and every subject, amounting to a total size greater than 2 GB.

**STEP-II-** The pre-possessing of the raw WESAD data files has been performed to remove the unwanted redundant values, noise and missing values from the dataset. Data integration is also performed to combine the individual data files of 15 users to attain the overall accuracy of the collected data set.

**STEP-III-** The class imbalance problem is resolved using the SMOTE (Synthetic Minority Oversampling Technique) and it provides the new balanced dataset in form of output.

**STEP-IV-** The newly obtained balanced dataset is considered as input to the well-established traditional machine learning classifiers for selection of the best suited hyper parameters across each model for the betterment of accuracy using the 10-fold class validation followed by 70-30 ratio policy.

**STEP-V -** As a result, it is visualized that Random forest classification models attain the best accuracy using the balanced dataset with a well-tuned set of hyper parameters.

**STEP-VI-** Correlation coefficient based Recursive feature elimination method (CoC-RFE) is applied for the selection of the optimal subset of features. For this purpose we have taken the samples of 10, 20, 30, 40 and 50 features.

**STEP-VII-** The statistical parameters (i.e., accuracy, f1-score, recall, precision,) have been computed for performance evaluation of the CoC-RFE- GB model.

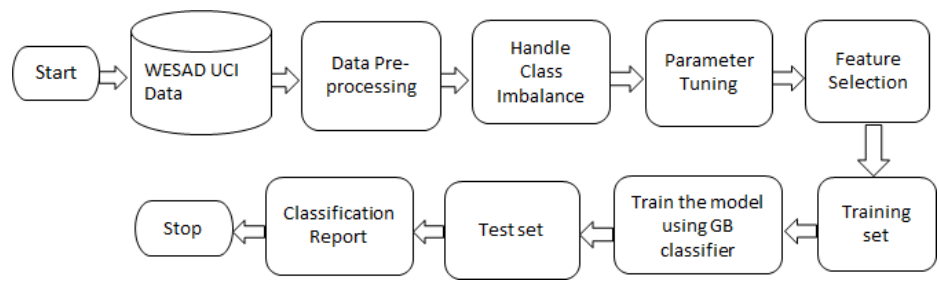

**Fig 3. Setup of proposed model**

### 4. Result and Discussion

Selecting the ideal model for an IoT-based smart healthcare system is a time-consuming and inventive process. The primary goal of this research is to develop a new hybrid model, CoC-REF-GB (Correlation Coefficient-based Recursive Feature Elimination Gradient Boosting Classifier), that can address the class imbalance problem, overfitting and underfitting data problems, and optimal feature selection problem with increased accuracy and precision. It will take the lead in automating the process of stress detection for the human well-being by using IoT-based data. This study makes use of a variety of known machine learning classifiers, including LR, SVC, KNN, LDA, RF, GB and a proposed CoC-RFE-GB model which has been shown in figure 4.

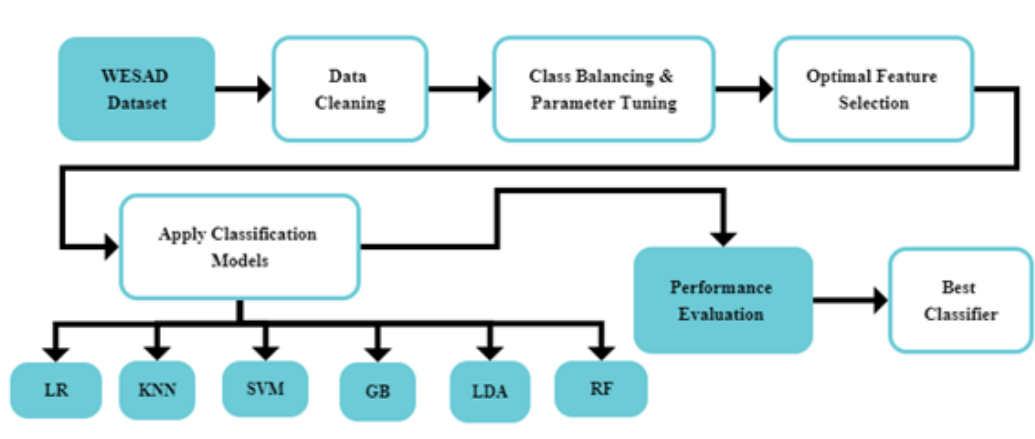

Fig 4. A quick look of various classification models

To choose the most acceptable machine learning classifier for the proposed model, a thorough comparison of six well-known models was conducted utilising a variety of performance assessment metrics that verify the classification result. The training - testing ratio used for this experiment is 70 -30. Table 1 summarises the statistical findings achieved throughout the experiment utilising the entire unbalanced dataset (processed dataset), the balanced dataset, and the hyper tuned balanced dataset. The experimental findings indicate that the newly proposed model achieves the maximum accuracy over the whole experiment when compared to the other conventional machine learning classifiers on a variety of statistical metrics (i.e. accuracy, f1- score, recall, precision).

Table 1. Comparison of various statistical measures between imbalanced, balanced and hyper tuned balanced dataset

| S. No. | Classification model | Statistical measures | Complete Imbalanced dataset | Balanced dataset | Hyper tuned balanced dataset |
|---|---|---|---|---|---|
| 1 | RF  | Accuracy  | 89.41 | 90.07 | 91.72 |
|   |     | F1-score  | 0.89  | 0.91  | 0.91  |
|   |     | Precision | 0.9   | 0.9   | 0.92  |
|   |     | Recall    | 0.89  | 0.9   | 0.91  |
| 2 | LR  | Accuracy  | 72.97 | 86.94 | 88.22 |
|   |     | F1-score  | 0.71  | 0.86  | 0.88  |
|   |     | Precision | 0.71  | 0.87  | 0.88  |
|   |     | Recall    | 0.72  | 0.86  | 0.88  |
| 3 | SVC | Accuracy  | 52.33 | 58.92 | 68.02 |
|   |     | F1-score  | 0.52  | 0.58  | 0.68  |
|   |     | Precision | 0.52  | 0.58  | 0.67  |
|   |     | Recall    | 0.52  | 0.58  | 0.68  |
| 4 | KNN | Accuracy  | 53.69 | 70.16 | 85.09 |
|   |     | F1-score  | 0.83  | 0.7   | 0.85  |
|   |     | Precision | 0.54  | 0.7   | 0.84  |
|   |     | Recall    | 0.54  | 0.7   | 0.85  |
| 5 | LDA | Accuracy  | 75.28 | 86.94 | 86.78 |
|   |     | F1-score  | 0.75  | 0.87  | 0.86  |
|   |     | Precision | 0.76  | 0.87  | 0.87  |
|   |     | Recall    | 0.75  | 0.86  | 0.86  |
| 6 | GB  | Accuracy  | 86.36 | 90.07 | 90.88 |
|   |     | F1-score  | 0.86  | 0.9   | 0.9   |
|   |     | Precision | 0.85  | 0.9   | 0.91  |
|   |     | Recall    | 0.86  | 0.9   | 0.9   |

Figure 5 shows the visualization of classification results which represents accuracy comparison using imbalanced dataset, balanced dataset and hyper tuned balanced dataset respectively. The comparative analysis shows GB models give highest accuracy with hyper tuned balanced dataset.

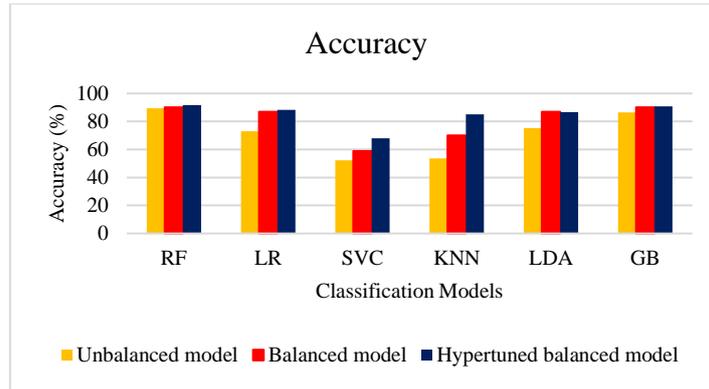

Fig 5. Accuracy comparison among various ML classifiers

Figure 6 shows f1- score comparison using imbalanced dataset, balanced dataset and hyper tuned balanced dataset respectively. The comparative analysis shows models give the highest f1- score value with hyper tuned balanced dataset.

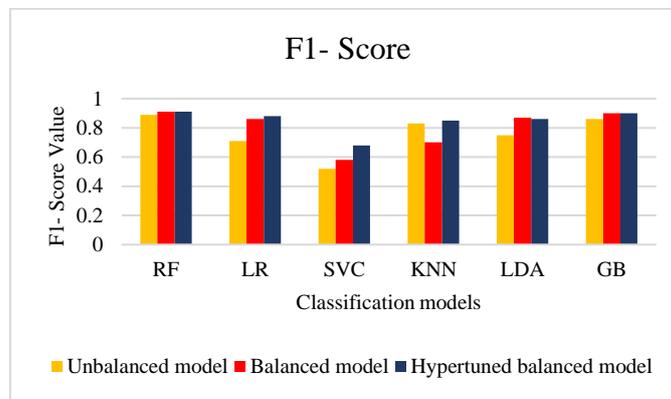

Fig 6. F1- Score comparison among various ML classifiers

Figure 7 performs comparison of precision results using imbalanced dataset, balanced dataset and hyper tuned balanced dataset respectively. The comparative analysis shows GB models give the highest precision value with hyper tuned balanced dataset.

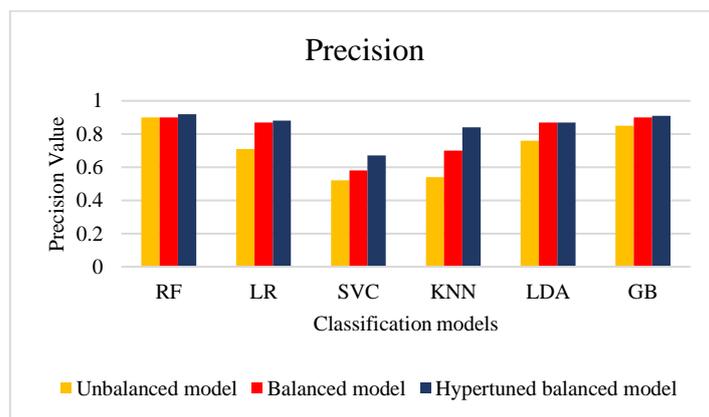

Fig 7. F1- Score comparison among various ML classifiers

Figure 8 performs comparison among recall values using imbalanced dataset, balanced dataset and hyper tuned balanced dataset respectively. The comparative analysis shows GB models give the highest recall value with hyper tuned balanced dataset.

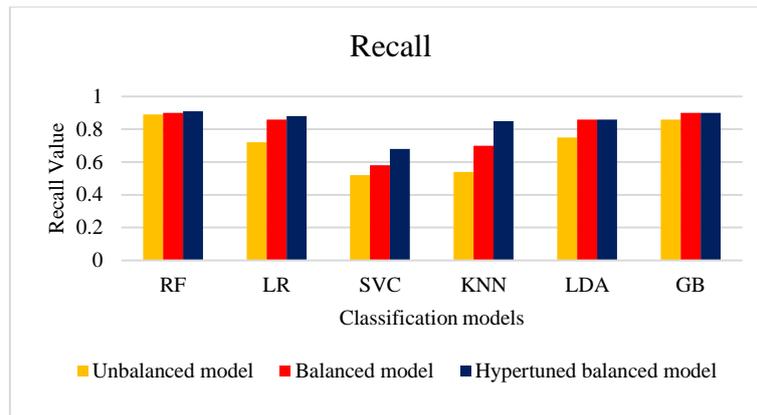

Fig 8. Recall comparison among various ML classifiers

**4.1 Accuracy of various classifier using Feature selection Techniques**

Table 2 shows the performance of four well established feature selection techniques i.e. ANOVA f- test, Mutual Information (MI), Recursive Feature Elimination (REF), and Correlation coefficient used for finding the best subset of features to obtain best accuracy of the model. The performance of these existing feature selection techniques were compared to the newly designed CoC-RFE feature selection algorithm. The highest accuracy achieved from this novel hybrid model CoC-RFE feature selection method clubbed with GB classifier.

Table 2: Performance evaluation of various feature selection techniques

| Feature selection method | No. of features selected | Accuracy | Optimal selected features |
|---|---|---|---|
| ANOVA f- test | 10 | 80.21 | 40 |
| | 20 | 82.30 | |
| | 30 | 86.47 | |
| | 40 | 90.75 | |
| | 50 | 88.26 | |
| Feature Importance | 10 | 85.71 | 40 |
| | 20 | 87.11 | |
| | 30 | 88.79 | |
| | 40 | 92.43 | |
| | 50 | 92.43 | |
| RFE | 10 | 76.8 | 50 |
| | 20 | 76.8 | |
| | 30 | 79.9 | |
| | 40 | 82.24 | |
| | 50 | 85.1 | |
| Correlation coefficient | 10 | 82.19 | 40 |
| | 20 | 85.37 | |
| | 30 | 86.15 | |
| | 40 | 93.13 | |
| | 50 | 90.24 | |
| CoC-RFE | 10 | 94.70 | 40 |
| | 20 | 95.51 | |
| | 30 | 96.15 | |
| | 40 | 98.56 | |
| | 50 | 98.56 | |

Figure 9 represents the graph for accuracy achieved using the various feature selection methods applied with the GB classification model. It is visualized that the newly developed CoC-RFE hybrid feature selection approach attains the highest accuracy in comparison with the other techniques.

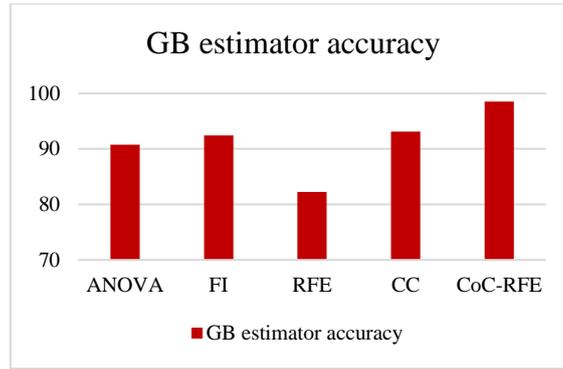

Fig 9. Comparative analysis of GB estimator with various Feature selection techniques

The classification accuracy of individual subjects is also computed using this novel approach CoC-RFE-GB classifier, which is represented in table 3.

Table 3: Classification report of individual user

| Subject ID | Accuracy GB (%) | F1-score | Precision | Recall |
|---|---|---|---|---|
| S2 | 95.65 % | 0.9552042 | 0.9444444 | 0.97222222 |
| S3 | 91.67 % | 0.9285714 | 0.9555555 | 0.91666666 |
| S4 | 95.65 % | 0.9543042 | 0.9465444 | 0.9622222 |
| S5 | 100% | 1.0 | 1.0 | 1.0 |
| S6 | 100% | 1.0 | 1.0 | 1.0 |
| S7 | 91.67 % | 0.9267399 | 0.9166666 | 0.95238095 |
| S8 | 100% | 1.0 | 1.0 | 1.0 |
| S9 | 100% | 1.0 | 1.0 | 1.0 |
| S10 | 100% | 1.0 | 1.0 | 1.0 |
| S11 | 100% | 1.0 | 1.0 | 1.0 |
| S13 | 100% | 1.0 | 1.0 | 1.0 |
| S14 | 95.83 % | 0.9496296 | 0.9333333 | 0.97435897 |
| S15 | 100% | 1.0 | 1.0 | 1.0 |
| S16 | 100% | 1.0 | 1.0 | 1.0 |
| S17 | 100% | 1.0 | 1.0 | 1.0 |

## 5. Conclusion and Future Scope

Lifestyle diseases (for example, stress and related complications) are one of the greatest worldwide threats to the human population. As a result, it is critical to monitor several physiological characteristics continuously and in real time. Developing an appropriate model for smart healthcare services is a difficult but fascinating undertaking. Various uses of smart healthcare systems include monitoring blood pressure (BP), glucose levels, oxygen levels, skin conductance and resistance, and electrocardiography (ECG). This article offers a novel model for stress lifestyle disease classification that is capable of resolving class imbalance issues and overcoming overfitting and underfitting issues via the use of fine-tuned sets of hyper parameters. Additionally, by picking the ideal subset of features, the model will lower the computational cost and complexity. The experimental findings indicate that the proposed model outperforms existing well-established machine learning classifiers in terms of accuracy across a variety of statistical metrics. Additionally, this research may be expanded from an algorithmic and data analytics perspective.